\documentclass[conference]{IEEEtran}
\usepackage{amsmath,amssymb}
\usepackage{graphicx}
\usepackage{mathrsfs}
\usepackage{tikz}
\usepackage{booktabs}
\usepackage{floatflt}
\usepackage{enumerate}
\usepackage{psfrag}	
\usepackage{array}
\usepackage{multirow,hhline}
\usepackage{exscale}
\usepackage{color}
\usepackage{colortbl}
\usepackage{epsfig,subfigure}
\usepackage{cite}
\usepackage{amsthm}

\renewcommand{\vec}[1]{\ensuremath{\mathbf{#1}}}
\newtheorem{theorem}{Theorem}
\newtheorem{lemma}{Lemma}
\newtheorem{remark}{Remark}
\newtheorem{corollary}{Corollary}

\begin{document}
\IEEEoverridecommandlockouts
\title{Signal Space Alignment for the Gaussian Y-Channel}
\author{
\IEEEauthorblockN{Anas Chaaban and Aydin Sezgin}
\IEEEauthorblockA{Chair of Communication Systems\\
RUB, 44780 Bochum, Germany\\
Email: {anas.chaaban@rub.de, aydin.sezgin@rub.de}}
}

\maketitle


\begin{abstract}
A multi-way communication network with three nodes and a relay is considered. The three nodes in this so-called Y-channel, communicate with each other in a bi-directional manner via the relay. Studying this setup is important due to its being an important milestone for characterizing the capacity of larger networks. A transmit strategy for the Gaussian Y-channel is proposed, which mimics a previously considered scheme for the deterministic approximation of the Y-channel. Namely, a scheme which uses nested-lattice codes and lattice alignment is used, to perform network coding. A new mode of operation is introduced, named `cyclic communication', which interestingly turns out to be an important component for achieving the capacity region of the Gaussian Y-channel within a constant gap.
\end{abstract}

\section{Introduction}
Multi-way communications was first studied by Shannon in \cite{Shannon_TWC} where the so-called two-way channel was considered. This setup consists of two nodes which act as transmitters and receivers in the same time, and its capacity is not known in general. By combining relaying and multi-way communications, we obtain the so-called multi-way relay channel. For instance, the two-way relay channel (or the bi-directional relay channel) consists of two nodes communicating with each other in both directions, via a relay. This setup was introduced in \cite{RankovWittneben} and later studied in \cite{KimDevroyeMitranTarokh,GunduzTuncelNayak,AvestimehrSezginTse,NarayananPravinSprintson} leading to an approximate characterization of the capacity region of the Gaussian case.

The multi-way relay channel with more nodes was also studied in \cite{GunduzYenerGoldsmithPoor} in a multicast scenario. In \cite{OngKellettJohnson}, the common-rate capacity of the Gaussian multi-way relay channel, where each user multi-casts a message to all other users, was obtained by using the so-called `functional decode-and-forward'. A broadcast variant of this multi-way relaying setup, the so called Y-channel, was considered in \cite{LeeLim}. Each user in the Y-channel sends two independent messages, one to each other user. \cite{LeeLim} considered the multiple-input multiple-output Y-channel. Namely, 3 MIMO nodes communicate via a MIMO relay. A transmission scheme exploiting signal space alignment \cite{MaddahAliMotahariKhandani_XChannel,CadambeJafar_KUserIC} was proposed, and its corresponding achievable degrees of freedom were calculated. In \cite{LeeLimChun}, it was shown that if the relay has more than $\lceil{3M/2}\rceil$ antennas where $M$ is the number of antennas at the other nodes, then the cut-set bound is asymptotically achievable, thus characterizing the degrees of freedom of the MIMO Y-channel under this condition.

We consider the single antenna Gaussian Y-channel. This case is not covered in \cite{LeeLimChun}, and as it turns out, the statement in \cite{LeeLimChun} does not apply here. In fact, it was shown in \cite{ChaabanSezginAvestimehr_Asilomar} that further bounds (other than the cut-set bounds) are required to characterize the degrees of freedom of the single antenna Y-channel. Thus, in the single antenna case, the cut-set bounds are not asymptotically achievable. From this point of view, it is worth to study the capacity of the SISO Y-channel as a separate problem.

In this paper, we propose a transmission scheme for the Gaussian Y-channel which utilizes nested-lattice codes in a functional decode-and-forward fashion, and derive its achievable rate region. It turns out that this scheme achieves the capacity region of the Y-channel within a constant gap.
To this end, the system model is given in section \ref{Sec:NotationAndSystemModel}. A toy example illustrating our scheme for the deterministic Y-channel is given in Section \ref{Sec:DYCAchievability}. The transmit strategy for the Gaussian Y-channel is described in Section \ref{Sec:GYCInnerBound} and we conclude with section \ref{Sec:Discussion}.

\section{System Model}
\label{Sec:NotationAndSystemModel}
\begin{figure}[t]
\centering
\includegraphics[width=0.8\columnwidth]{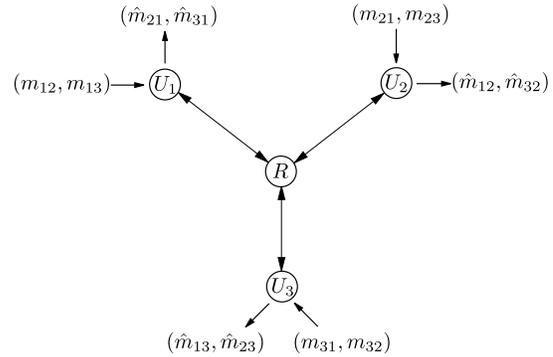}
\caption{The Y-channel showing incoming and outgoing messages.}
\label{Fig:Model}
\end{figure}

The Y-channel is the multi-way relaying setup shown in Fig. \ref{Fig:Model}. Each user $U_k$ sends a message to each other user via the relay. 
A code for the Y-channel, an achievable rate tuple $\vec{R}=(R_{12},R_{13},R_{21},R_{23},R_{31},R_{32})$, and the 6-dimensional capacity region is defined in the classical information theoretic sense (see \cite{ChaabanSezginISIT,ChaabanSezginAvestimehr_Asilomar}). In our Gaussian Y-channel (GYC), the variables are real valued. The relay receives
\begin{align*}
y_{ri}=h_1x_{1i}+h_2x_{2i}+h_3x_{3i}+z_{ri},
\end{align*}
in time instant $i$, where $z_{ri}$ is a realization of an independent and identically distributed Gaussian noise with zero mean and unit variance (i.i.d. $\mathcal{N}(0,1)$) and $h_1,h_2,h_3\in\mathbb{R}$ are the channel coefficients from the users to the relay. Without loss of generality, we assume that $h_1^2\geq h_2^2\geq h_3^2$. The received signal at user $j$ is given by
\begin{align*}
y_{ji}=h_jx_{ri}+z_{ji},
\end{align*}
where $x_{ri}$ is the relay signal at time instant $i$, and $z_{ji}$ is a realization of an i.i.d. $\mathcal{N}(0,1)$ noise. The channel are assumed to be reciprocal, and all nodes have a power constraint $P$, i.e., $\frac{1}{n}\sum_{i=1}^n\mathbb{E}[X_{ri}^2]\leq P$, and $\frac{1}{n}\sum_{i=1}^n\mathbb{E}[X_{ji}^2]\leq P$. Here, $n$ is the length of the code. To illustrate our achievable scheme for the GYC, we start by considering a toy example for the linear-shift deterministic \cite{AvestimehrDiggaviTse} Y-channel (DYC) defined in \cite{ChaabanSezginISIT}.

\section{A Capacity Achieving Scheme for the DYC}
\label{Sec:DYCAchievability}
In this section, we describe briefly the network coding based scheme in \cite{ChaabanSezginISIT} by considering the following toy example. In the DYC, we distinguish between three different patterns of information flow as follows:
\begin{itemize}
\item[$b$)] \textbf{B}i-directional: where $R_{jk}$ and $R_{kj}$ are both non-zero for some  $j,k\in\{1,2,3\}$, $j\neq k$. 
\item[$c$)] \textbf{C}yclic: where $R_{jk}$, $R_{kl}$, and $R_{lj}$ are non-zero while $R_{kj}=R_{lk}=R_{jl}=0$ for distinct $j,k,l\in\{1,2,3\}$.
\item[$u$)]\textbf{U}ni-directional: where neither case b) nor c) holds. 
\end{itemize}

\subsection{DYC: A Toy Example}
\begin{figure}[t]
\centering
\includegraphics[width=0.62\columnwidth]{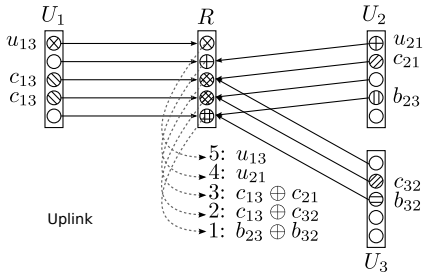}
\caption{The DYC of the toy example with an illustration of our transmit strategy in the uplink. The circles denote bit positions, and the arrows denote bit pipes.}
\label{DYC543U}
\end{figure}

Consider the DYC shown in Fig. \ref{DYC543U}. The received signal at the relay is given here by the $\mod 2$ sum of the bits arriving at each level. Let us choose the following rate tuple $\vec{R}=(0,2,2,1,0,2)$, and see how our scheme achieves this rate tuple. It can easily be checked, that the schemes used in the bi-directional relay channel \cite{AvestimehrSezginTse} (only cases $b$ and $u$ above) do not suffice to achieve this rate tuple.

\begin{figure}[t]
\centering
\includegraphics[width=\columnwidth]{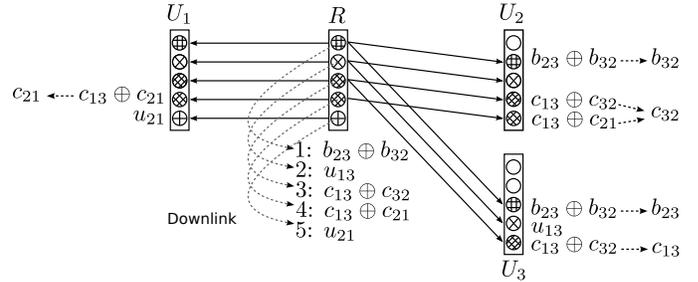}
\caption{The DYC of the toy example with an illustration of our transmit strategy in the downlink.}
\label{DYC543D}
\end{figure}

We write $\vec{R}=\vec{R}^b+\vec{R}^c+\vec{R}^u$, where $\vec{R}^b=(0,0,0,1,0,1)$, $\vec{R}^c=(0,1,1,0,0,1)$, and $\vec{R}^u=(0,1,1,0,0,0)$. Notice that $\vec{R}^b$ resembles bi-directional information flow between $U_2$ and $U_3$ with a rate of 1 bit per channel use in each direction. To achieve this rate tuple, let $U_2$ send one bit $b_{23}$ on relay level 1 in the uplink, and let $U_3$ also send 1 bit $b_{32}$ on the same level (Fig. \ref{DYC543U}). Thus, the relay receives $b_{23}\oplus b_{32}$ on level 1.
The relay then forwards $b_{23}\oplus b_{32}$ on the highest level in the downlink (Fig. \ref{DYC543D}). Upon receiving $b_{23}\oplus b_{32}$, $U_2$ and $U_3$ are able to extract their desired bits, $b_{32}$ and $b_{23}$, respectively, which achieves $\vec{R}^b$. We call this strategy the bi-directional strategy.

The rate tuple $\vec{R}^c$ resembles cyclic information flow, where $U_1$, $U_2$, and $U_3$ want to send 1 bit each $c_{13}$, $c_{21}$, and $c_{32}$ to $U_3$, $U_1$ and $U_2$, respectively, thus forming the cycle $1\to3\to2\to1$. Here, we use a cyclic strategy as follows. Let $U_1$ send $c_{13}$ on both relay levels 2 and 3, $U_2$ send $c_{21}$ on relay level 3, and $U_3$ send $c_{32}$ on relay level 2. The relay thus receives $c_{13}\oplus c_{32}$ and $c_{13}\oplus c_{21}$ on levels 2 and 3, respectively (Fig. \ref{DYC543U}). It then forwards these sums on levels 3 and 4 (\ref{DYC543D}). Each receiver receives $c_{13}\oplus c_{21}$ and $c_{13}\oplus c_{32}$, and by adding them up, it can construct $c_{32}\oplus c_{21}$. Then, given its transmitted bit, each receiver is able to calculate the other two bits of the cyclic information flow, which achieves $\vec{R}^c$.
%

Finally, $\vec{R}^u$ can be easily achieved using a uni-directional strategy. Here, $U_1$ and $U_2$ send one bit each, $u_{13}$ and $u_{21}$, to levels 5 and 4 at the relay, respectively (Fig. \ref{DYC543U}). The relay forwards these bits on levels 2 and 5, respectively, and users $U_1$ and $U_3$ are then able to recover both desired bits. This achieves $\vec{R}^u$ and consequently, we have achieved the rate tuple $\vec{R}$. 

The given scheme consumed all the levels at the relay to achieve $\vec{R}$ (see Fig. \ref{DYC543U} and \ref{DYC543D}). If we replace the cyclic strategy, which uses 2 levels at the relay for communicating 3 bits, by the uni-directional strategy, then we do not leave enough levels free to achieve $\vec{R}^u$. This shows the importance of the cyclic strategy.
Finally, we note that it was shown in \cite{ChaabanSezginISIT} that the given scheme achieves the capacity region of the DYC. In the next section, we extend this scheme to the Gaussian case, with the aid of nested-lattice codes.

\section{The GYC: An Achievable Scheme}
\label{Sec:GYCInnerBound}
We adapt the scheme in Section \ref{Sec:DYCAchievability} to the Gaussian case. Namely, we utilize network coding realized with lattice codes \cite{Loeliger} to mimic the DYC scheme. We start with a brief introduction on lattice codes, before proceeding to describe the achievable scheme.

\subsection{Nested-lattice codes}
A lattice $\Lambda$ with $n$-dimensions is a subset of $\mathbb{R}^n$, where $\lambda_1,\lambda_2\in\Lambda\Rightarrow\lambda_1+\lambda_2\in\Lambda$. The fundamental Voronoi of $\Lambda$, $\mathcal{V}(\Lambda)$, is the Voronoi region around the origin. Nested-lattice codes are constructed using two lattices, a coarse lattice $\Lambda_c$ and a fine lattice $\Lambda_f$ where $\Lambda_c\subset\Lambda_f$. We denote a nested-lattice code by the pair $(\Lambda_f,\Lambda_c)$, where the codewords are chosen as the points $\lambda_f\in\Lambda_f\cap\mathcal{V}(\Lambda_c)$. The power and the rate of such code is defined by $\Lambda_c$ and by the size of the set $\Lambda_f\cap\mathcal{V}(\Lambda_c)$, respectively. In the sequel, we are going to need the following result from \cite{NarayananPravinSprintson}. 

Assume that two nodes A and B, with messages $m_A$ and $m_B$, respectively, want to exchange these messages via a relay node. The two nodes use the same nested-lattice codebook $(\Lambda_f,\Lambda_c)$ with power $P$, and rate $R$ to encode their messages to $x_k^n=(\lambda_k-d_k)\mod\Lambda_c$, $k\in\{A,B\}$, where $\lambda_A,\lambda_B\in\Lambda_f\cap\mathcal{V}(\Lambda_c)$, $d_A$ and $d_B$ are $n$-dimensional dither vectors uniformely distributed over $\mathcal{V}(\Lambda_c)$ \cite{NazerGastpar}, known at all nodes. The relay receives $y_R^n=x_A^n+x_B^n+z_R^n$ where $z_R^n$ is an i.i.d. $\mathcal{N}(0,\sigma^2)$ noise. Let $C(x)=(1/2)\log(1+x)$, and $C^+(x)=\max\{0,C(x)\}$.
\begin{lemma}[\cite{NarayananPravinSprintson}]
\label{Lemma:AlignDecodeRate}
The relay can decode the sum $(\lambda_A+\lambda_B)\mod \Lambda_c$ from $y_R^n$ reliably as long as $R\leq C^+\left(\frac{P}{\sigma^2}-\frac{1}{2}\right)$. Moreover, node A knowing $(\lambda_A+\lambda_B)\mod \Lambda_c$ and $\lambda_A$ can extract $\lambda_B$ and hence also $m_B$.
\end{lemma}

\subsection{Uplink}
Now, we proceed with describing the transmission scheme. In the uplink, $U_i$ splits each message $m_{ij}$ into three parts:
\begin{itemize}
\item a bi-directional message $m_{ij}^b$ with rate $R_{ij}^b$, 
\item a cyclic message $m_{ij}^c$ with rate $R_{ij}^c$, and
\item a uni-directional message $m_{ij}^u$ with rate $R_{ij}^u$.
\end{itemize}
Thus, we have $R_{ij}=R_{ij}^b+R_{ij}^c+R_{ij}^u$. The messages $m_{ij}^b$, $m_{ij}^c$ and $m_{ij}^u$ are communicated using a bi-directional, a cyclic, and a uni-directional strategy, respectively. The rates of the messages satisfy $R_{12}^b=R_{21}^b$, $R_{13}^b=R_{31}^b$, $R_{23}^b=R_{32}^b$, $R_{12}^c=R_{23}^c=R_{31}^c\triangleq R_{123}^c$, $R_{13}^c=R_{32}^c=R_{21}^c\triangleq R_{132}^c$.

\subsubsection{Encoding bi-directional messages}
\label{Sec:BiDirStreams}
The users use nested-lattices to encode the bi-directional messages. Let us consider the bi-directional communication between users 1 and 2, i.e., the messages $m_{12}^b$ and $m_{21}^b$. $U_2$ uses a nested-lattice code $(\Lambda_{21}^b,\Lambda_{21,c}^b)$. The rate of the code is $R_{21}^b$ and the power is $P_{21}^b$. The message $m_{21}^b$ is mapped into $\lambda_{21}^b$. $U_1$ uses a scaled version of $(\Lambda_{21}^b,\Lambda_{21,c}^b)$ to encode $m_{12}^b$ such that the bi-directional signals align at the relay. That is, $U_1$ uses $(\Lambda_{12}^b,\Lambda_{12,c}^b)=\frac{h_2}{h_1}\left(\Lambda_{21}^b,\Lambda_{21,c}^b\right)$, for encoding $m_{12}^b$ to $\lambda_{12}^b$. Using this encoding, the rate of the nested-lattice code $(\Lambda_{12}^b,\Lambda_{12,c}^b)$ is $R_{12}^b=R_{21}^b$, its power $P_{12}^b$ satisfies 
\begin{align}
\label{A12bVsA21b}
h_1^2P_{12}^b=h_2^2P_{21}^b,
\end{align}
and $(h_1\lambda_{12}^b+h_2\lambda_{21}^b)\mod h_2\Lambda_{21,c}^b\in h_2\Lambda_{21}^b\cap\mathcal{V}(h_2\Lambda_{21,c}^b)$ which is a useful property as we shall see in Section \ref{DecAtRelay}. Then, $U_1$ and $U_2$ construct the signals $b_{12}^n$ and $b_{21}^n$ as follows,
\begin{align*}
b_{ij}^n&=(\lambda_{ij}^b-d_{ij}^b)\mod \Lambda_{ij,c}^b
\end{align*}
with $i\neq j$, $i,j\in\{1,2\}$, where $d_{ij}^b$ is a random dither, uniformly distributed over $\mathcal{V}(\Lambda_{ij,c}^b)$, known at all nodes (see \cite{NarayananPravinSprintson}). Similarly, $m_{31}^b$ and $m_{13}^b$ are encoded into $b_{31}^n$ and $b_{13}^n$ with powers $P_{31}^b$ and $P_{31}^b$, respectively, and $m_{32}^b$ and $m_{23}^b$ into $b_{32}^n$ and $b_{23}^n$ with powers $P_{32}^b$ and $P_{23}^b$, where
\begin{align}
\label{A13bVsA31b}
h_1^2P_{13}^b=h_3^2P_{31}^b,\quad h_2^2P_{23}^b=h_3^2P_{32}^b.
\end{align}

\subsubsection{Encoding cyclic messages}
Consider $m_{12}^c$, $m_{23}^c$ and $m_{31}^c$ (all with rate $R_{123}^c$) constituting the cycle $1\to2\to3\to1$. To communicate these messages, $U_1$ and $U_3$ encode $m_{12}^c$ and $m_{31}^c$ to $c_{12}^n=(\lambda_{12}^c-d_{12}^c)\mod \Lambda_{12,c}^c$ and $c_{31}^n=(\lambda_{31}^c-d_{31}^c)\mod \Lambda_{31,c}^c$ using nested-lattice codes $(\Lambda_{12}^c,\Lambda_{12,c}^c)$ and $(\Lambda_{31}^c,\Lambda_{31,c}^c)$ with powers $P_{12}^c$ and $P_{31}^c$, respectively. 

Now, $U_2$ sends $m_{23}^c$ encoded in two different signals: one signal aligned with $\lambda_{12}^c$, and one signal aligned with $\lambda_{31}^c$. This mimics the scheme used for the cyclic messages in the the DYC (Section \ref{Sec:DYCAchievability}). Alignment is guaranteed using the nested-lattice construction, in a similar way as for the bi-directional messages (Section \ref{Sec:BiDirStreams}). Namely, $U_2$ maps $m_{23}^c$ to $c_{23}^n=(\lambda_{23}^c-d_{23}^c)\mod \Lambda_{23,c}^c$ and $\tilde{c}_{23}^n=(\tilde{\lambda}_{23}^c-\tilde{d}_{23}^c)\mod \tilde{\Lambda}_{23,c}^c$, using nested-lattice codes $(\Lambda_{23}^c,\Lambda_{23,c}^c)=\frac{h_1}{h_2}(\Lambda_{12}^c,\Lambda_{12,c}^c)$ and $(\tilde{\Lambda}_{23}^c,\tilde{\Lambda}_{23,c}^c)=\frac{h_3}{h_2}(\Lambda_{31}^c,\Lambda_{31,c}^c)$ with powers $P_{23}^c$ and $\tilde{P}_{23}^c$, respectively, such that 
\begin{align}
\label{A12cVsA23c}
h_1^2P_{12}^c&=h_2^2P_{23}^c,\quad h_2^2\tilde{P}_{23}^c=h_3^2P_{31}^c.
\end{align} 
Notice that this ensures alignment of the codes $(\Lambda_{23}^c,\Lambda_{23,c}^c)$ and $(\Lambda_{12}^c,\Lambda_{12,c}^c)$, as well as  $(\Lambda_{31}^c,\Lambda_{31,c}^c)$ and  $(\tilde{\Lambda}_{23}^c,\tilde{\Lambda}_{23,c}^c)$ at the relay, allowing the relay to decode $(h_1\lambda_{12}^c+h_2\lambda_{23}^c)\mod h_2\Lambda_{23,c}^c$ and $(h_2\tilde{\lambda}_{23}^c+h_3\lambda_{31}^c)\mod h_3\Lambda_{31,c}^c$ as we shall see in Section \ref{DecAtRelay}. The messages of the other cycle $1\to3\to2\to1$ are encoded similarly, to $c_{13}^n=(\lambda_{13}^c-d_{13}^c)\mod \Lambda_{13,c}^c$, $\tilde{c}_{13}^c=(\tilde{\lambda}_{13}^c-\tilde{d}_{13}^c)\mod\tilde{\Lambda}_{13,c}^c$, $c_{32}^n=(\lambda_{32}^c-d_{32}^c)\mod \Lambda_{32,c}^c$, $c_{21}^n=(\lambda_{21}^c-d_{21}^c)\mod \Lambda_{21,c}^c$ with powers $P_{13}^c$, $\tilde{P}_{13}^c$, $P_{32}^c$, and $P_{21}^c$, respectively, such that $(\Lambda_{13}^c,\Lambda_{13,c}^c)=\frac{h_3}{h_1}(\Lambda_{32}^c,\Lambda_{32,c}^c)$, $(\tilde{\Lambda}_{13}^c,\tilde{\Lambda}_{13,c}^c)=\frac{h_2}{h_1}(\Lambda_{21}^c,\Lambda_{21,c}^c)$, 
\begin{align}
\label{A13cVsA32c}
h_1^2P_{13}^c&=h_3^2P_{32}^c,\quad h_1^2\tilde{P}_{13}^c=h_2^2P_{21}^c.
\end{align}

\subsubsection{Encoding uni-directional messages}
The uni-directional message $m_{ij}^u$ with rate $R_{ij}^u$ is encoded using a Gaussian code. Namely, $m_{ij}^u$ is mapped $u_{ij}^n$, an i.i.d. $\mathcal{N}(0,P_{ij}^u)$ sequence.

\subsubsection{Transmit signals}
Each user then transmits the superposition of all its codewords. For instance, $U_1$ sends
\begin{align*}
x_1^n&=b_{12}^n+b_{13}^n+c_{12}^n+c_{13}^n+\tilde{c}_{13}^n+u_{12}^n+u_{13}^n.
\end{align*}
Since each node has a power constraint $P$, then we must have
\begin{align}
\label{Cond:Alpha1}
P_{12}^b+P_{13}^b+P_{12}^c+P_{13}^c+\tilde{P}_{13}^c+P_{12}^u+P_{13}^u&=P_1\leq P\\
\label{Cond:Alpha2}
P_{21}^b+P_{23}^b+P_{21}^c+P_{23}^c+\tilde{P}_{23}^c+P_{21}^u+P_{23}^u&=P_2\leq P\\
\label{Cond:Alpha3}
P_{31}^b+P_{32}^b+P_{31}^c+P_{32}^c+P_{31}^u+P_{32}^u&=P_3\leq P,
\end{align}
Next, we describe the decoding process at the relay.

\subsubsection{Decoding at the relay}
\label{DecAtRelay}
The relay decodes the uni-directional signals $u_{12}^n$, $u_{13}^n$, $u_{21}^n$, and $u_{23}^n$ first, successively in the given order while treating all the remaining signals as noise. The effective noise power while decoding $u_{12}^n$ is given by $h_3^2P_3+h_2^2P_2+h_1^2(P_1-P_{12}^u)+1$. Then, reliable decoding of $u_{12}^n$ is possible under the rate constraint 
\begin{align*}
R_{12}^u&\leq C\left(\frac{h_1^2P_{12}^u}{1+h_3^2P_3+h_2^2P_2+h_1^2(P-P_{12}^u)}\right).
\end{align*}

\begin{figure*}
\begin{align}
\label{R21uU}
R_{21}^u&\leq C\left(\frac{h_2^2P_{21}^u}{1+h_3^2P_3+h_2^2(P_2-P_{21}^u)+h_1^2(P-P_{12}^u-P_{13}^u)}\right),\\
\label{R23uU}
R_{23}^u&\leq C\left(\frac{h_2^2P_{23}^u}{1+h_3^2P_3+h_2^2(P_2-P_{21}^u-P_{23}^u)+h_1^2(P-P_{12}^u-P_{13}^u)}\right).
\end{align}
\hrule
\end{figure*}

After decoding $u_{12}^n$ and subtracting its contribution from the received signal at the relay, the other signals $u_{13}^n$, $u_{21}^n$, and $u_{23}^n$ are decoded. Reliable decoding is possible if \eqref{R21uU}-\eqref{R13uU} are satisfied.
\begin{align}
\label{R13uU}
R_{13}^u&\leq C\left(\frac{h_1^2P_{13}^u}{1+h_3^2P_3+h_2^2P_2+h_1^2(P-P_{12}^u-P_{13}^u)}\right).
\end{align}
Using \eqref{A12bVsA21b}-\eqref{A13cVsA32c}, we can write the remaining noise variance as $\sigma^2+2h_2^2(P_{21}^b+P_{23}^c+P_{21}^c)$ where $\sigma^2=1+h_3^2(2P_{31}^b+2P_{32}^b+2P_{31}^c+2P_{32}^c+P_{31}^u+P_{32}^u)$. Next, the relay decodes the superposition $(h_2\lambda_{21}^c+h_1\tilde{\lambda}_{13}^c)\mod h_2\Lambda_{21,c}^c$ (which is possible since this quantity belongs to the nested lattice code $(h_2\Lambda_{21}^c,h_2\Lambda_{21,c}^c)$), then $(h_1\lambda_{12}^c+h_2\lambda_{23}^c)\mod h_2\Lambda_{23,c}^c$ afterwards and then $(h_1\lambda_{12}^b+h_2\lambda_{21}^b)\mod h_2\Lambda_{21,c}^b$ successively in this order using successive compute-and-forward \cite{Nazer_IZS2012} while treating the remaining interference as noise. From Lemma \ref{Lemma:AlignDecodeRate}, the decoding of these signals is possible reliably as long as 
\begin{align*}
R_{132}^c&\leq C^+\left(\frac{h_2^2P_{21}^c}{\sigma^2+2h_2^2(P_{21}^b+P_{23}^c)}-\frac{1}{2}\right),\\
R_{123}^c&\leq C^+\left(\frac{h_2^2P_{23}^c}{\sigma^2+2h_2^2P_{21}^b}-\frac{1}{2}\right)\\
R_{21}^b&\leq C^+\left(\frac{h_2^2P_{21}^b}{\sigma^2}-\frac{1}{2}\right).
\end{align*}

Next, the uni-directional signals $u_{31}^n$ and $u_{32}^n$ are decoded, then the superposition of the cyclic signals $(h_1\lambda_{13}^c+h_3\lambda_{32}^c)\mod h_3\Lambda_{32,c}^c$ and $(h_2\tilde{\lambda}_{23}^c+h_3\lambda_{31}^c)\mod h_3\Lambda_{31,c}^c$, and finally, the superposition of the bi-directional signals $(h_1\lambda_{13}^b+h_3\lambda_{31}^b)\mod h_3\Lambda_{31,c}^b$ and $(h_2\lambda_{23}^b+h_3\lambda_{32}^b)\mod h_3\Lambda_{32,c}^b$, successively in the given order (again using successive compute-and-forward \cite{Nazer_IZS2012}), resulting in the following rate constraints
\begin{align*}
R_{31}^u&\leq C\left(\frac{h_3^2P_{31}^u}{1+h_3^2(2P_{32}^b+2P_{31}^b+2P_{31}^c+2P_{32}^c+P_{32}^u)}\right)\\
R_{32}^u&\leq C\left(\frac{h_3^2P_{32}^u}{1+2h_3^2(P_{32}^b+P_{31}^b+P_{31}^c+P_{32}^c)}\right)\\
R_{132}^c&\leq C^+\left(\frac{h_3^2P_{32}^c}{1+2h_3^2(P_{32}^b+P_{31}^b+P_{31}^c)}-\frac{1}{2}\right)\\
R_{123}^c&\leq C^+\left(\frac{h_3^2P_{31}^c}{1+2h_3^2(P_{32}^b+P_{31}^b)}-\frac{1}{2}\right)\\
R_{31}^b&\leq C^+\left(\frac{h_3^2P_{31}^b}{1+2h_3^2P_{32}^b}-\frac{1}{2}\right),\quad 
R_{32}^b\leq C^+\left(h_3^2P_{32}^b-\frac{1}{2}\right).
\end{align*}

\subsection{Downlink}
In the downlink, the relay maps each of the decoded signals into an index which is then encoded into a Gaussian codeword as follows:\\ 
$u_{ij}^n\to l_{ij}^u \to t_{ij}^n$,\\ 
$(h_1\lambda_{12}^c+h_2\lambda_{23}^c)\mod h_2\Lambda_{23,c}^c\to l_{12}^c\to s_{12}^n$,\\ 
$(h_2\tilde{\lambda}_{23}^c+h_3\lambda_{31}^c)\mod h_3\Lambda_{31,c}^c\to l_{31}^c\to  s_{31}^n$,\\ 
$(h_2\lambda_{21}^c+h_1\tilde{\lambda}_{13}^c)\mod h_2\Lambda_{21,c}^c\to l_{21}^c\to  s_{21}^n$,\\
$(h_1\lambda_{13}^c+h_3\lambda_{32}^c)\mod h_3\Lambda_{32,c}^c\to l_{32}^c\to  s_{32}^n$,\\
$(h_1\lambda_{12}^b+h_2\lambda_{21}^b)\mod h_2\Lambda_{21,c}^b\to l_{21}^b\to  r_{21}^n$,\\
$(h_1\lambda_{13}^b+h_3\lambda_{31}^b)\mod h_3\Lambda_{31,c}^b\to l_{31}^b\to  r_{31}^n$,\\
$(h_2\lambda_{23}^b+h_3\lambda_{32}^b)\mod h_3\Lambda_{32,c}^b\to l_{32}^b\to  r_{32}^n$. 

The relay allocates a power $P_{r,ij}^u$ to $t_{ij}^n$, i.e., $t_{ij}^n$ is i.i.d $\mathcal{N}(0,P_{r,ij}^u)$. It also allocates $P_{r,ij}^c$ to $s_{ij}^n$ and $P_{r,ij}^b$ to $r_{ij}^n$. For the power constraint to be satisfied, it is required that the sum of these powers fulfils
\begin{align}
\label{Cond:Beta}
\sum P_{r,ij}^u+\sum P_{r,ij}^c+\sum P_{r,ij}^b\leq P. 
\end{align}
The relay then sends the superposition of all $t_{ij}^n$, $s_{ij}^n$, and $r_{ij}^n$, denoted $x_r^n$. The decoding process at each of the nodes $U_1$, $U_2$, and $U_3$ is described next.

\subsubsection{Decoding at $U_3$}
$U_3$ decodes the messages $l_{13}^u$, $l_{23}^u$, $l_{31}^c$, $l_{32}^c$, $l_{31}^b$, $l_{32}^b$ in this order while treating the other signals as noise. The necessary rate constraints for reliable decoding are
\begin{align*}
R_{13}^u&\leq C\left(\frac{h_3^2P_{r,13}^u}{\sigma_{r1}^2+h_3^2(P_{r,23}^u+P_{r,32}^c+P_{r,31}^c+P_{r,31}^b+P_{r,32}^b)}\right)\\
R_{23}^u&\leq C\left(\frac{h_3^2P_{r,23}^u}{\sigma_{r1}^2+h_3^2(P_{r,32}^c+P_{r,31}^c+P_{r,31}^b+P_{r,32}^b)}\right)\\
R_{132}^c&\leq C\left(\frac{h_3^2P_{r,32}^c}{\sigma_{r1}^2+h_3^2(P_{r,31}^c+P_{r,31}^b+P_{r,32}^b)}\right)\\
R_{123}^c&\leq C\left(\frac{h_3^2P_{r,31}^c}{\sigma_{r1}^2+h_3^2(P_{r,31}^b+P_{r,32}^b)}\right)\\
R_{31}^b&\leq C\left(\frac{h_3^2P_{r,31}^b}{\sigma_{r1}^2+h_3^2P_{r,32}^b}\right),\quad R_{32}^b\leq C\left(\frac{h_3^2P_{r,32}^b}{\sigma_{r1}^2}\right)
\end{align*}
where $\sigma_{r1}^2=1+h_3^2(P_{r,12}^u+P_{r,32}^u+P_{r,12}^c+P_{r,21}^c+P_{r,21}^b+P_{r,21}^u+P_{r,31}^u)$. By decoding $l_{13}^u$ and $l_{23}^u$, the third user can obtain the uni-directional messages $m_{13}^u$ and $m_{23}^u$. By decoding $l_{32}^c$, the third user can obtain the superposition $(h_1\lambda_{13}^c+h_3\lambda_{32}^c)\mod h_3\Lambda_{32,c}^c$. Knowing $\lambda_{32}^c$, $U_3$ can extract $\lambda_{13}^c$ and hence obtain the desired cyclic communication message $m_{13}^c$ (cf. Lemma \ref{Lemma:AlignDecodeRate}). Similarly, by decoding $l_{31}^c$, $l_{31}^b$ and $l_{32}^b$, the messages $m_{23}^c$, $m_{13}^b$, and $m_{23}^b$ can be obtained. Notice that $U_3$ can remove $t_{31}^^n$ and $t_{32}^^n$ before decoding. We do not remove them for the purpose of having more unified expressions for all receivers.
%

\subsubsection{Decoding at $U_2$}
\label{DecAtU2}
Since $U_3$ can decode its desired messages, $U_2$ can also decode $U_3$'s desired messages, since $h_2^2\geq h_3^2$. After decoding the messages intended to $U_3$, $U_2$ decodes the messages $l_{12}^u$, $l_{32}^u$, $l_{12}^c$, $l_{21}^c$, and $l_{21}^b$ successively in this order while treating the remaining signals as noise. The following rate constraints have to be fulfilled
\begin{align*}
R_{12}^u&\leq C\left(\frac{h_2^2P_{r,12}^u}{\sigma_{r2}^2+h_2^2(P_{r,32}^u+P_{r,12}^c+P_{r,21}^c+P_{r,21}^b)}\right)\\
R_{32}^u&\leq C\left(\frac{h_2^2P_{r,32}^u}{\sigma_{r2}^2+h_2^2(P_{r,12}^c+P_{r,21}^c+P_{r,21}^b)}\right)\\
R_{123}^c&\leq C\left(\frac{h_2^2P_{r,12}^c}{\sigma_{r2}^2+h_2^2(P_{r,21}^c+P_{r,21}^b)}\right)\\
R_{132}^c&\leq C\left(\frac{h_2^2P_{r,21}^c}{\sigma_{r2}^2+h_2^2P_{r,21}^b}\right), \quad R_{21}^b\leq C\left(\frac{h_2^2P_{r,21}^b}{\sigma_{r2}^2}\right)
\end{align*}
where $\sigma_{r2}^2=1+h_2^2(P_{r,21}^u+P_{r,31}^u)$. In this way, $U_2$ is able to obtain $m_{12}^u$, $m_{32}^u$, $m_{12}^c$, $m_{13}^c$, $m_{12}^b$ and $m_{32}^b$. Notice that $m_{13}^c$ is not desired by $U_2$, but it can be used in combination with $l_{32}^c$ (recall that this can be decoded by $U_2$ since it can be decoded by $U_3$) to obtain $m_{32}^c$ which is a desired message.

\subsubsection{Decoding at $U_1$}
\label{DecAtU1}
Finally, $U_1$ decodes all signals that are decodable by $U_2$ and $U_3$, followed by $l_{21}^u$ and $l_{31}^u$ with the following rate constraints
\begin{align*}
R_{21}^u&\leq C\left(\frac{h_1^2P_{r,21}^u}{1+h_1^2P_{r,31}^u}\right),\quad R_{31}^u\leq C\left(h_1^2P_{r,31}^u\right).
\end{align*}

This allows $U_1$ to obtain all its desired messages. Let the region achieved by this scheme, for a given power allocation satisfying the power constraints, be denoted $\mathcal{R}_g(P_{ij}^u,\ P_{ij}^c,\ P_{ij}^b,P_{r,ij}^u,\ P_{r,ij}^c,\ P_{r,ij}^b)$.
Then we have the following inner bound.
\begin{theorem}
The union over all possible power allocations satisfying the rate constraints \eqref{Cond:Alpha1}-\eqref{Cond:Alpha3}, and \eqref{Cond:Beta} of the region $\mathcal{R}_g(P_{ij}^u,\ P_{ij}^c,\ P_{ij}^b,P_{r,ij}^u,\ P_{r,ij}^c,\ P_{r,ij}^b)$ is an inner bound on the capacity region $\mathcal{C}_g$ of the GYC.
\end{theorem}

\begin{remark}
Notice that a larger inner bound can be achieved if we remove $t_{31}^n$ and $t_{32}^n$ before decoding at $U_3$, and $t_{21}^n$ before decoding at $U_2$. Moreover, all the nodes can use different decoding orders to enlarge the inner bound. We do not consider these possibilities in this paper due to lack of space, however, the given scheme is sufficient for the main result of the paper given next.
\end{remark}

The given scheme achieves, within a constant gap of 7/6 per dimension, the capacity region of the GYC. Namely, the following region is achievable.

\begin{corollary}
For the given GYC, the region $\underline{\mathcal{C}}'_g$ given by 
\begin{align*}
R_{31}+R_{32}&\leq C(h_3^2P)-2\\
R_{13}+R_{23}&\leq C(h_3^2P)-2\\
R_{12}+R_{13}+R_{32}&\leq C(h_2^2P+h_3^2P)-3\\
R_{13}+R_{23}+R_{12}&\leq C(h_2^2P+h_3^2P)-3\\
R_{12}+R_{31}+R_{32}&\leq C(h_1^2P+h_2^2P)-3\\
R_{13}+R_{23}+R_{21}&\leq C(h_1^2P+h_3^2P)-3\\
R_{21}+R_{31}+R_{23}&\leq C((|h_2|+|h_3|)^2P)-7/2\\
R_{21}+R_{31}+R_{32}&\leq C((|h_2|+|h_3|)^2P)-7/2,
\end{align*}
is achievable.
\end{corollary}

The region $\underline{\mathcal{C}}'_g$ is within a constant gap of an outer bound on the capacity region of the GYG (bounds given in \cite{ChaabanSezginAvestimehr_Asilomar}). Details are not given due to the lack of space. 

\section{Conclusion}
\label{Sec:Discussion}
A transmission scheme is proposed for the Y-channel by using network coding ideas. The achievability scheme is based on three different strategies, a bi-directional, a cyclic, and a uni-directional strategy. While the first and the last are used to establish the capacity of the bi-directional relay channel, the second is new. Nested-lattices have been used to establish network coding. The achievable rate region of the given scheme is given. It turns out that the given scheme achieves the capacity region within a constant gap.

\bibliography{myBib}

\begin{thebibliography}{10}
\providecommand{\url}[1]{#1}
\csname url@samestyle\endcsname
\providecommand{\newblock}{\relax}
\providecommand{\bibinfo}[2]{#2}
\providecommand{\BIBentrySTDinterwordspacing}{\spaceskip=0pt\relax}
\providecommand{\BIBentryALTinterwordstretchfactor}{4}
\providecommand{\BIBentryALTinterwordspacing}{\spaceskip=\fontdimen2\font plus
\BIBentryALTinterwordstretchfactor\fontdimen3\font minus
  \fontdimen4\font\relax}
\providecommand{\BIBforeignlanguage}[2]{{%
\expandafter\ifx\csname l@#1\endcsname\relax
\typeout{** WARNING: IEEEtran.bst: No hyphenation pattern has been}%
\typeout{** loaded for the language `#1'. Using the pattern for}%
\typeout{** the default language instead.}%
\else
\language=\csname l@#1\endcsname
\fi
#2}}
\providecommand{\BIBdecl}{\relax}
\BIBdecl

\bibitem{Shannon_TWC}
C.~Shannon, ``{Two-way communication channels},'' in \emph{Proc. of Fourth
  Berkeley Symposium on Mathematics, Statistics, and Probability}, vol.~1,
  1961, pp. 611--644.

\bibitem{RankovWittneben}
B.~Rankov and A.~Wittneben, ``{Spectral efficient signaling for half-duplex
  relay channels},'' in \emph{Proc. of the Asilomar Conference on Signals,
  Systems, and Computers}, Pacific Grove, CA, Nov. 2005.

\bibitem{KimDevroyeMitranTarokh}
S.~Kim, N.~Devroye, P.~Mitran, and V.~Tarokh, ``{Comparisons of bi-directional
  relaying protocols},'' in \emph{Proc. of the IEEE Sarnoff Symposium},
  Princeton, NJ, Apr. 2008.

\bibitem{GunduzTuncelNayak}
D.~G\"und\"uz, E.~Tuncel, and J.~Nayak, ``{Rate regions for the separated
  two-way relay channel},'' in \emph{Proc. of the 46th Annual Allerton
  Conference on Communication, Control, and Computing}, Urbana-Champaign, IL,
  Sep. 2008, pp. 1333 -- 1340.

\bibitem{AvestimehrSezginTse}
A.~S. Avestimehr, A.Sezgin, and D.~Tse, ``{Capacity of the two-way relay
  channel within a constant gap},'' \emph{European Trans. in
  Telecommunications}, 2009.

\bibitem{NarayananPravinSprintson}
K.~Narayanan, M.~P. Wilson, and A.~Sprintson, ``{Joint physical layer coding
  and network coding for bi-directional relaying},'' in \emph{Proc. of the
  Forty-Fifth Allerton Conference}, Illinois, USA, Sep. 2007.

\bibitem{GunduzYenerGoldsmithPoor}
D.~G\"und\"uz, A.~Yener, A.~Goldsmith, and H.~V. Poor, ``{The multi-way relay
  channel},'' in \emph{Proc. of IEEE ISIT}, Seoul, South Korea, Jun. 2009.

\bibitem{OngKellettJohnson}
L.~Ong, C.~Kellett, and S.~Johnson, ``{Capacity theorems for the AWGN multi-way
  relay channel},'' in \emph{Proc. of IEEE ISIT}, 2010.

\bibitem{LeeLim}
N.~Lee and J.-B. Lim, ``{A novel signaling for communication on MIMO Y channel:
  Signal space alignment for network coding},'' in \emph{Proc. of IEEE ISIT},
  vol.~1, Seoul, Jun. 2009, pp. 2892 -- 2896.

\bibitem{MaddahAliMotahariKhandani_XChannel}
M.~Maddah-Ali, A.~Motahari, and A.~Khandani, ``{Communication over MIMO X
  channels: interference alignment, decomposition, and performance analysis},''
  \emph{IEEE Trans. on Info. Theory}, vol.~54, no.~8, pp. 3457--3470, Aug.
  2008.

\bibitem{CadambeJafar_KUserIC}
V.~R. Cadambe and S.~A. Jafar, ``{Interference alignment and the degrees of
  freedom for the K user interference channel},'' \emph{IEEE Trans. on Info.
  Theory}, vol.~54, no.~8, pp. 3425--3441, Aug. 2008.

\bibitem{LeeLimChun}
N.~Lee, J.-B. Lim, and J.~Chun, ``{Degrees of freedom of the MIMO Y channel:
  Signal space alignment for network coding},'' \emph{IEEE Trans. on Info.
  Theory}, vol.~56, no.~7, pp. 3332--3342, Jul. 2010.

\bibitem{ChaabanSezginAvestimehr_Asilomar}
A.~Chaaban, A.~Sezgin, and A.~S. Avestimehr, ``{On the sum capacity of the
  Y-Channel},'' in \emph{Proc. of 42nd Asilomar Conference on Signals, Systems
  and Computers}, Pacific Grove, CA, USA, Nov. 2011.

\bibitem{ChaabanSezginISIT}
A.~Chaaban and A.~Sezgin, ``{The capacity region of the linear shift
  deterministic Y-channel},'' in \emph{IEEE International Symposium on Info.
  Theory (ISIT)}, St. Petersburg, July 31-Aug. 5 2011, pp. 2457 -- 2461.

\bibitem{AvestimehrDiggaviTse}
A.~S. Avestimehr, S.~Diggavi, and D.~Tse, ``{A deterministic approach to
  wireless relay networks},'' in \emph{Proc. of Allerton Conference}, 2007.

\bibitem{Loeliger}
H.~A. Loeliger, ``{Averaging bounds for lattices and linear codes},''
  \emph{IEEE Trans. on Info. Theory}, vol.~43, no.~6, p. 1767–1773, Nov.
  1997.

\bibitem{NazerGastpar}
B.~Nazer and M.~Gastpar, ``{Compute-and-Forward: Harnessing interference
  through structured codes},'' \emph{IEEE Trans. on Info. Theory}, vol.~57,
  no.~10, pp. 6463 -- 6486, Oct. 2011.

\bibitem{Nazer_IZS2012}
B.~Nazer, ``{Successive compute-and-forward},'' in \emph{Proc. of the 22nd
  International Zurich Seminar on Communication (IZS 2012)}, Zurich,
  Switzerland, March 2012.

\end{thebibliography}

\end{document}